\documentclass[sigconf, authorversion]{acmart}
\usepackage{algorithmic,algorithm}
\usepackage{caption}
\usepackage{tabularx}
\usepackage{graphics}
\usepackage{multirow}
\usepackage{colortbl}
\usepackage{xcolor,pifont}
\newcommand*\colourcheck[1]{%
  \expandafter\newcommand\csname #1check\endcsname{\textcolor{#1}{\ding{52}}}%
}
\colourcheck{blue}
\colourcheck{green}
\colourcheck{red}
\newcommand*\colourcross[1]{%
  \expandafter\newcommand\csname #1cross\endcsname{\textcolor{#1}{\ding{55}}}%
}
\colourcross{red}
\usepackage[title]{appendix}

\AtBeginDocument{%
  \providecommand\BibTeX{{%
    \normalfont B\kern-0.5em{\scshape i\kern-0.25em b}\kern-0.8em\TeX}}}

\copyrightyear{2023}
\acmYear{2023}
\setcopyright{rightsretained}
\acmConference[CUI '23]{ACM conference on Conversational User Interfaces}{July 19--21, 2023}{Eindhoven, Netherlands}
\acmBooktitle{ACM conference on Conversational User Interfaces (CUI '23), July 19--21, 2023, Eindhoven, Netherlands}
\acmDOI{10.1145/3571884.3604308}
\acmISBN{979-8-4007-0014-9/23/07}

\begin{document}

\title{Democratizing Chatbot Debugging: A Computational Framework for Evaluating and Explaining Inappropriate Chatbot Responses}

\author{Xu Han}
\affiliation{%
  \institution{University of Colorado Boulder}
  \city{Boulder}
  \state{CO}
  \country{USA}}
\email{xuha2442@colorado.edu}

\author{Michelle Zhou}
\affiliation{%
  \institution{Juji, Inc.}
  \city{San Jose}
  \state{CA}
  \country{USA}}
\email{mzhou@acm.org}

\author{Yichen Wang}
\affiliation{%
  \institution{University of Colorado Boulder}
  \city{Boulder}
  \state{CO}
  \country{USA}}
\email{yichen.wang@colorado.edu}

\author{Wenxi Chen}
\affiliation{%
  \institution{Juji, Inc.}
  \city{San Jose}
  \state{CA}
  \country{USA}}
\email{wchen@juji-inc.com}

\author{Tom Yeh}

\affiliation{%
  \institution{University of Colorado Boulder}
  \city{Boulder}
  \state{CO}
  \country{USA}}
\email{tom.yeh@colorado.edu}

\renewcommand{\shorttitle}{Democratizing Chatbot Debugging}


\begin{abstract}
Evaluating and understanding the inappropriateness of chatbot behaviors can be challenging, particularly for chatbot designers without technical backgrounds. To democratize the debugging process of chatbot misbehaviors for non-technical designers, we propose a framework that leverages dialogue act (DA) modeling to automate the evaluation and explanation of chatbot response inappropriateness. The framework first produces characterizations of context-aware DAs based on discourse analysis theory and real-world human-chatbot transcripts. It then automatically extracts features to identify the appropriateness level of a response and can explain the causes of the inappropriate response by examining the DA mismatch between the response and its conversational context.  Using interview chatbots as a testbed, our framework achieves comparable classification accuracy with higher explainability and fewer computational resources than the deep learning baseline, making it the first step in utilizing DAs for chatbot response appropriateness evaluation and explanation. 

\end{abstract}



\begin{teaserfigure}
     \centering
         \includegraphics[width=0.8\textwidth, height=0.23\textheight]{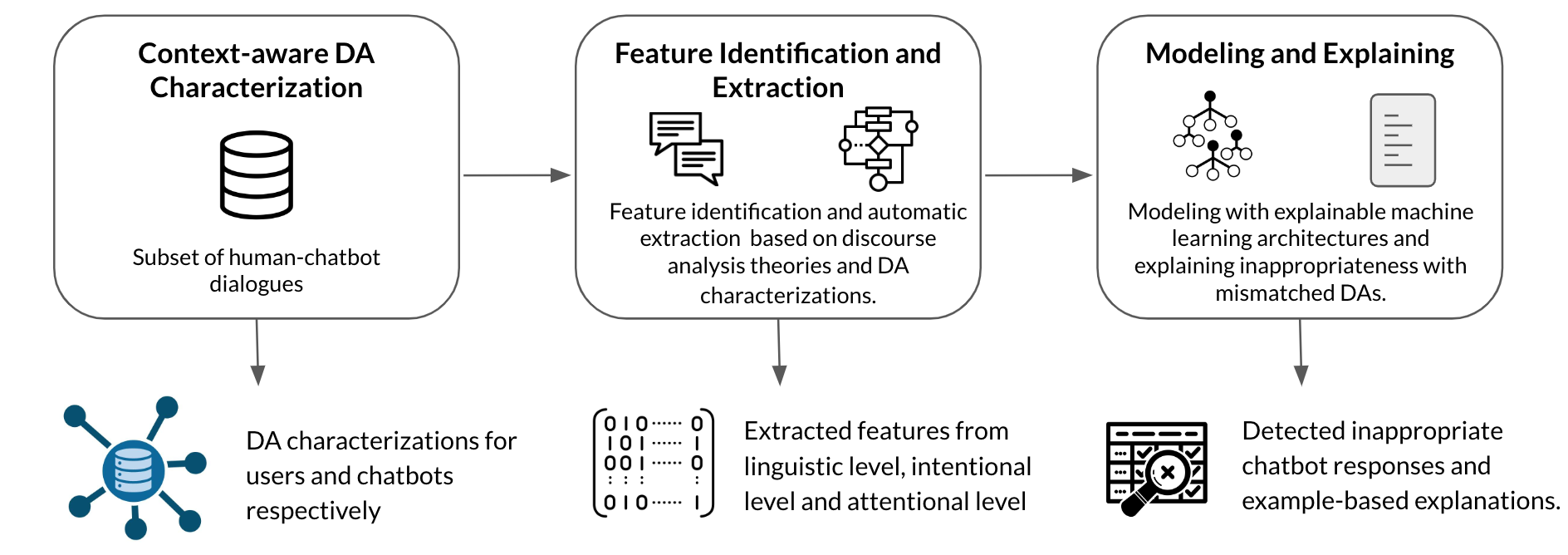}
         \caption{The three-step modeling workflow for our proposed computational framework for Evaluating and Explaining Inappropriate Chatbot Responses}
         \Description{This picture describes the three-step modeling workflow for our proposed computational framework for Evaluating and Explaining Inappropriate Chatbot Responses. It is composed of three blocks, representing the step of "context-aware DA characterization", "feature identification and extraction" and "modeling and explaining" respectively.}
      \label{fig:framework}
      \vspace{0.8cm}
\end{teaserfigure}

\maketitle

\section{Introduction}
\begin{figure*}
     \centering
     \begin{minipage}[b]{0.45\textwidth}
         \centering
         \includegraphics[width=\textwidth]{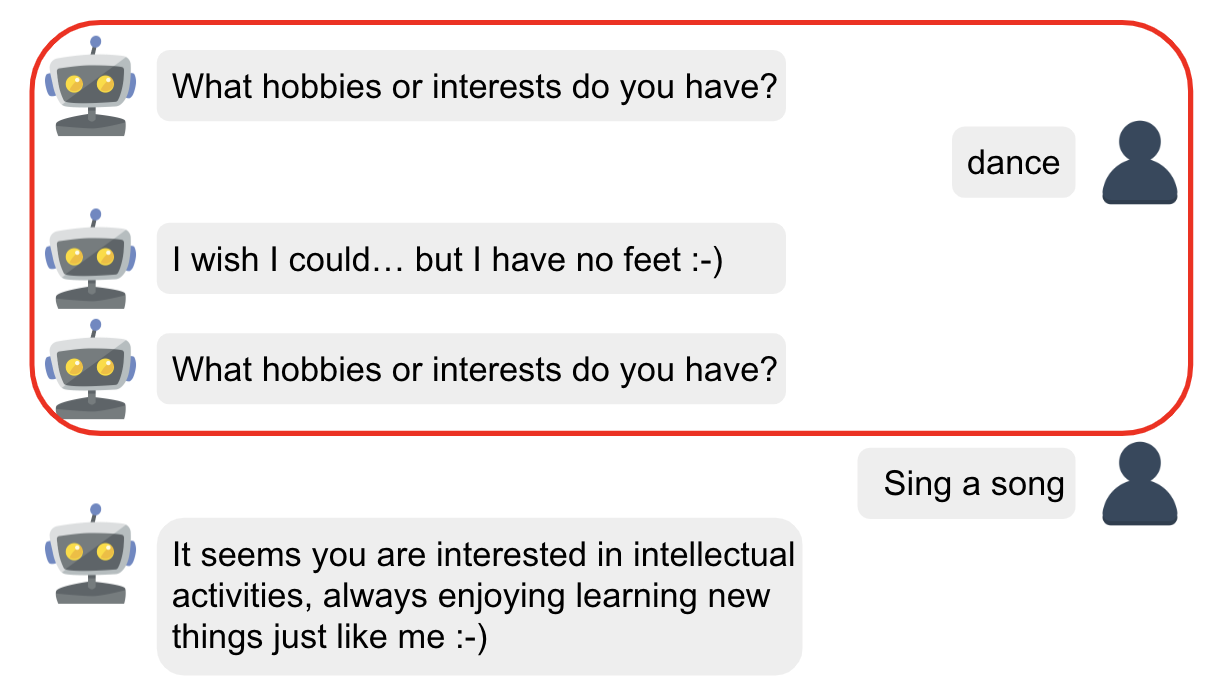}
         \caption*{(a)}
         \Description{This example human-chatbot dialogue shows an interaction where the chatbot incorrectly recognizes the user’s input as a request and responds inappropriately}
         \vspace{-0.2cm}
     \end{minipage}
     \hfill
     \begin{minipage}[b]{0.45\textwidth}
         \centering
         \includegraphics[width=\textwidth]{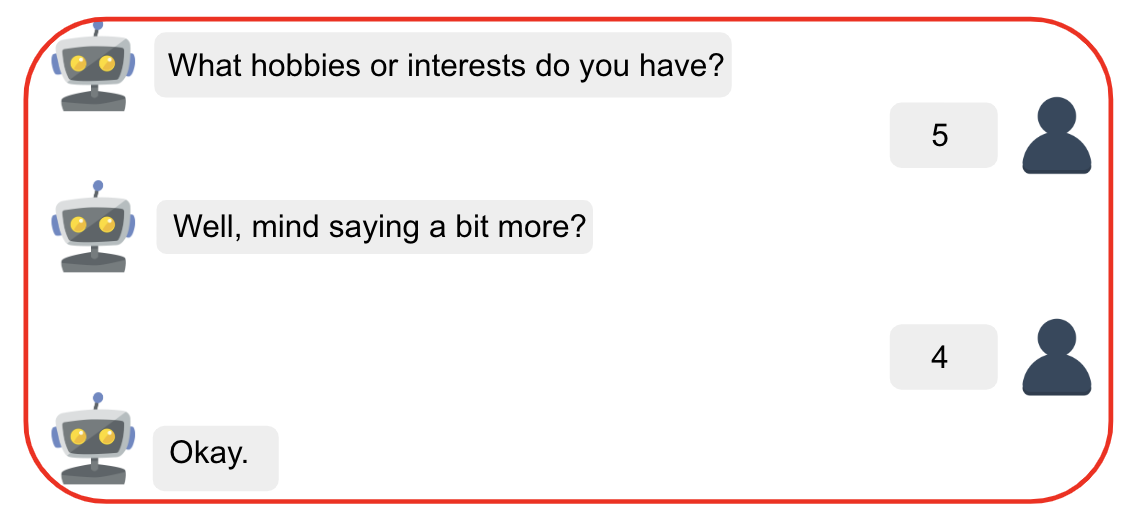}
         \caption*{(b)}
         \Description{This example human-chatbot dialogue shows an interaction where the chatbot mistakes the user's gibberish input for a legitimate answer and accepts it with an acknowledgment.}
         \vspace{-0.2cm}
     \end{minipage}
     \hfill
     \caption{Example human-chatbot dialogues. (a) The chatbot incorrectly recognizes the user’s input ("\textit{dance}") as a request and responds accordingly; (b) the chatbot mistakes the user's gibberish input ("\textit{5}") for a legitimate answer and accepts it with an acknowledgment.}
      \label{fig:exampledialogue}
\end{figure*}

Chatbot systems like ChatGPT\cite{chatgpt} engage users in one-on-one text-based conversations by responding to user inputs correspondingly. While natural language generation (NLG) approaches, such as the use of large language models (LLMs), have made significant progress in generating syntactically well-formed chatbot responses \cite{welivita-pu-2020-taxonomy, bubeck2023sparks}, it remains challenging to ensure that these responses are appropriate for the given conversational contexts \cite{spring2019empathic}. 
The mismatch between the chatbot responses and the contexts can happen due to issues like context complexity\cite{raheja-tetreault-2019-dialogue}, limitations in NLG model architectures\cite{bubeck2023sparks}, and dataset bias\cite{10.1145/3466132.3466134}. For example, the chatbot response \textit{"I wish I could... but I have no feet :-)"} may be appropriate in the context of asking the chatbot to dance, but it's entirely inappropriate if the user indicates dancing as the hobby (Figure~\ref{fig:exampledialogue}(a)). Such inappropriate responses (i.e., chatbot misbehavior) can lead to poor user experience or even abandoned conversations \cite{han2021designing}. Therefore, it's critical for chatbot designers to ensure response appropriateness during the design processes.

In light of this, designers often conduct pilot studies to evaluate chatbot response appropriateness and iterate their designs accordingly (i.e., chatbot debugging) \cite{han2021designing}. However, designers without technical backgrounds may face two challenges when it comes to \textbf{detecting} and \textbf{understanding} potentially inappropriate responses revealed by these studies. First, it can be difficult for them to detect inappropriate responses without adequate computational resources. Examining all chat transcripts collected from pilot studies to locate inappropriate responses, such as the example in Figure~\ref{fig:exampledialogue}(a), is a laborious and time-consuming task to perform manually \cite{han2021designing}. Even if designers opt to develop an automatic model for inappropriate response detection, they may be limited by a lack of access to training data and required computing power. Second, even if non-technical designers are able to locate all inappropriate responses, it can still be difficult for them to understand why they occur and how to address them. Given the wide variety of conversational contexts, chatbots can exhibit very different types of inappropriateness. For example, in Figure~\ref{fig:exampledialogue}(a), the chatbot incorrectly recognizes the user's input (\textit{"dance"}) as a request and responds accordingly, whereas in Figure~\ref{fig:exampledialogue}(b), the chatbot mistakes the user's gibberish input for a legitimate answer and accepts it with an acknowledgment. This high degree of variability in chatbot inappropriateness can make it challenging for non-technical designers to understand and address them within a unified framework.

To democratize the debugging process of chatbot misbehaviors for non-technical designers, we propose a computational framework to evaluate and explain chatbot response inappropriateness through characterizing context-aware dialogue acts (DAs). 
Our framework draws inspiration from recent works that combine DA characterization and neural response generation tasks \cite{welivita-pu-2020-taxonomy,xu2018towards, montenegro2019dialogue, wu2021semantic}. These studies have shown the promise of utilizing DA modeling to enhance chatbot response quality, making them more controllable and interpretable. For example, Xu et al. \cite{xu2018towards} incorporate DAs as policies to improve their open-domain chatbot response generation model. With this in mind, our framework first guides the development of context-aware DA characterization of human-chatbot dialogues. Next, it identifies and extracts computational features based on the DA characterizations, and then trains automatic detection models to evaluate the appropriateness of a chatbot response. By utilizing DA characterization, our framework can explain the causes of inappropriate responses by examining the DA mismatch between the response and its conversational context. 

To the best of our knowledge, our framework is the first to incorporate DA characterization into the evaluation and explanation of chatbot inappropriate responses. To test the framework, we used \textit{interview chatbots} as a testbed and developed the first context-aware characterizations of DAs in human-interview chatbot interactions. It also achieved comparable accuracy in detecting inappropriate responses compared to the deep learning baseline, while offering greater explainability and requiring fewer computational resources.

\section{Methods}
\subsection{Testbed and Dataset}

\textbf{Testbed.} To ensure practicality, we have selected interview chatbots as our testbed, given their widespread use in a variety of applications, including research and job interviews \cite{Xiao2019-hk, Xiao2020-fr}. Interview chatbots utilize generative AI technology to engage users in text-based, one-on-one conversations, making them an ideal testbed for our study. Specifically, they are suitable for our study for several reasons: firstly, they support both task-oriented and social dialogues, making them representative of current chatbot systems; secondly, the dialogues between human and interview chatbots tend to follow a concise and controllable pattern of "\textit{interview question} (from chatbot) - \textit{answer} (from user) - \textit{response} (from chatbot)" \cite{han2021designing}, which facilitates our analysis of response appropriateness. Importantly, findings from interview chatbots can potentially be generalized to other chatbot categories \cite{han2021designing}.

\begin{table}[htb]
\centering
\renewcommand\arraystretch{1.2}
\caption{Interview Topics Used in Our Dataset}
\begin{tabular}{ll} 
      \hline
      Q1 & What hobbies or interests do you have?     \\
      Q2 & What do you do now for a living?   \\
      Q3 & What are your strongest qualities as a friend?     \\
      Q4 & \begin{tabular}[c]{@{}l@{}}{Tell me about a time when you didn’t know if you would} \\ {make it. How did you overcome that challenge?} \end{tabular} \\
      \hline
  \end{tabular}
~\label{tb:topictable}
\vspace{-0.2cm}
\end{table}

~\\
\noindent \textbf{Dataset.} We study real-world dialogues collected through the interview chatbots supported by Juji \footnote{https://juji.io/}, a publicly available chatbot platform where chatbot designers can create, customize, and deploy a chatbot with either a graphical user interface (GUI) or an interactive development environment (IDE) \cite{noauthor_Juji}.
We analyzed a dataset of 5342 real-world human-chatbot dialogues with 8987 chatbot responses in total, accumulated from various interview chatbots developed by Juji's designers, including personality survey bots \cite{fan_sun_liu_zhao_zhang_chen_glorioso_hack_2023}. These chatbots were active in the wild for dialogue transcript collection from February 2021 to July 2021. Each dialogue in the dataset was associated with one of the Juji built-in topics shown in Table~\ref{tb:topictable}. To ensure quality, we manually reviewed each dialogue, excluding those without any end-users inputs. The collection of these 5342 dialogues involves 2155 participants, most of whom are university students and their families with various backgrounds. For our study, we recruited two dialogue researchers to annotate all 8987 chatbot responses using three labels: \textit{Inappropriate}, \textit{Appropriate}, and \textit{Neutral}. Overall, the two annotators had achieved an inter-annotator agreement of 0.795 (Cohen's $\kappa$), which indicates a level of substantial agreement. When there were disagreements, the two annotators resolved the disagreement together through a discussion.

\begin{table*}
\renewcommand\arraystretch{1.8}
 \caption{Context-aware Characterization of User DAs When Interacting with Interview Chatbots}
\centering
\resizebox{2.0\columnwidth}{!}{\begin{tabular}{l!{\color{black}\vrule}ll}
\hline 
\multicolumn{1}{c}{\huge DA Categories} & \multicolumn{1}{c}{\huge Synopsis} & \multicolumn{1}{c}{\huge Typical Examples *} \\  
\hline
\multicolumn{1}{c}{\begin{tabular}[c]{@{}l@{}}\huge user-answer-relevant \end{tabular}} & \begin{tabular}[c]{@{}l@{}} \huge Giving relevant answers \\ \huge to the interview questions \end{tabular} & \begin{tabular}[c]{@{}l@{}}\huge \textit{"S: What things frighten you now?"} \\ \huge \textit{"U: \textbf{My future is the most terrifying.}"}\end{tabular} \\
\hline
\multicolumn{1}{c}{ \begin{tabular}[c]{@{}l@{}}\huge user-question-relevant \end{tabular} }& \begin{tabular}[c]{@{}l@{}} \huge Questioning for further details or \\\huge starting chitchat under the same topic \end{tabular} & \begin{tabular}[c]{@{}l@{}}\huge \textit{"S: What things frighten you now?"} \\ \huge \textit{"U:\textbf{ Why are you asking \textbf{?}}"}\end{tabular} \\
\hline
\multicolumn{1}{c}{\huge user-respond-irrelevant}  & \begin{tabular}[c]{@{}l@{}} \huge Responding with irrelevant information \\ \huge to the interview questions\end{tabular} & \begin{tabular}[c]{@{}l@{}}\huge \textit{"S: What hobbies or interests do you have?"} \\ \huge \textit{"U:\textbf{ I like blue most.}"}\end{tabular}\\
\hline
\multicolumn{1}{c}{ \begin{tabular}[c]{@{}l@{}}\huge user-question-irrelevant \end{tabular} }& \begin{tabular}[c]{@{}l@{}} \huge Questioning about different topics to start chitchat \end{tabular} & \begin{tabular}[c]{@{}l@{}}\huge \textit{"S: What hobbies or interests do you have?"} \\ \huge \textit{"U: \textbf{I like swimming. What are your capabilities?}"}\end{tabular} \\
\hline
\multicolumn{1}{c}{ \begin{tabular}[c]{@{}l@{}}\huge user-excuses \end{tabular} }& \begin{tabular}[c]{@{}l@{}} \huge Dodging answering the interview questions/digression \\ \huge with various excuses\end{tabular} & \begin{tabular}[c]{@{}l@{}}\huge \textit{"S: What were the worst parts of your childhood?"} \\ \huge \textit{"U:\textbf{ This is personal.}"}\end{tabular} \\
\hline
\multicolumn{1}{c}{ \begin{tabular}[c]{@{}l@{}}\huge user-acknowledge \end{tabular} }& \begin{tabular}[c]{@{}l@{}} \huge Accepting or admitting the chatbot's utterances \end{tabular} & \begin{tabular}[c]{@{}l@{}}\huge \textit{"U: How long is our chat gonna last?"} \\ \huge \textit{"S: If part of the chat progress bar is still red,}\\ \huge \textit{it indicates that our chat is still in progress.} \\ \huge \textit{ It will end before you know it."} \\ \huge \textit{"U: \textbf{Got it!}"}\end{tabular} \\
\hline
\multicolumn{1}{c}{ \begin{tabular}[c]{@{}l@{}}\huge user-request \end{tabular} }& \begin{tabular}[c]{@{}l@{}} \huge User's requests to the chatbot \end{tabular} & \begin{tabular}[c]{@{}l@{}} \huge \textit{"U:\textbf{ Tell me a joke.}"}\end{tabular} \\
\hline
\multicolumn{1}{c}{ \begin{tabular}[c]{@{}l@{}}\huge user-command \end{tabular} }& \begin{tabular}[c]{@{}l@{}} \huge User's commands on managing the chat-flow \end{tabular} & \begin{tabular}[c]{@{}l@{}} \huge \textit{"U:\textbf{ Next question.}"}\end{tabular} \\
\hline
\multicolumn{1}{c}{ \begin{tabular}[c]{@{}l@{}}\huge user-complain \end{tabular} }& \begin{tabular}[c]{@{}l@{}} \huge Complaining about the chatting experience or else \end{tabular} & \begin{tabular}[c]{@{}l@{}}\huge \textit{"S: What do you do now for a living?"} \\ \huge \textit{"U:\textbf{ You didn't listen. I just answered it.}"}\end{tabular}\\
\hline
\multicolumn{1}{c}{ \begin{tabular}[c]{@{}l@{}}\huge user-social-obligations \end{tabular} }& \begin{tabular}[c]{@{}l@{}} \huge Apology, greeting, thanking and etc. \end{tabular} & \begin{tabular}[c]{@{}l@{}}\huge \textit{"S: I hear you... would love to help} \\ \huge \textit{ when I have the power to do so."} \\ \huge \textit{"U:\textbf{ Thank you!}"}\end{tabular} \\
\hline
\multicolumn{1}{c}{ \begin{tabular}[c]{@{}l@{}}\huge user-gibberish \end{tabular} }& \begin{tabular}[c]{@{}l@{}} \huge user gives gibberish \end{tabular} & \begin{tabular}[c]{@{}l@{}}\huge \textit{"S: What hobbies or interests do you have?"} \\ \huge \textit{"U:\textbf{ blea blahe}"}\end{tabular} \\
\hline
\multicolumn{1}{c}{ \begin{tabular}[c]{@{}l@{}}\huge user-other \end{tabular} }& \begin{tabular}[c]{@{}l@{}} \huge Sentences do not belong to \\ \huge any of the categories above \end{tabular} & \begin{tabular}[c]{@{}l@{}}\huge \textit{"S: What hobbies or interests do you have?"} \\ \huge \textit{"U:\textbf{ Wow.}"}\end{tabular} \\
\hline
\multicolumn{2}{c}{ \begin{tabular}[c]{@{}l@{}}\huge * Note: "S" denotes the chatbot system while "U" denotes the user. \\ \huge Examples are for demonstration purposes only, not necessarily from the original transcripts.\end{tabular}}
\end{tabular}}
  
~\label{tb:usertaxonomy}
\end{table*}

\begin{table*}
\renewcommand\arraystretch{1.8}
\caption{Context-aware Characterization of Interview Chatbot DAs to Previous User Inputs}
\centering
\resizebox{2.0\columnwidth}{!}{\begin{tabular}{l!{\color{black}\vrule}ll}
\hline 
\multicolumn{1}{c}{\huge Chatbot Behavior Categories} & \multicolumn{1}{c}{\huge Synopsis} & \multicolumn{1}{c}{\huge Typical Examples *} \\  
\hline
\multicolumn{1}{c}{\begin{tabular}[c]{@{}l@{}}\huge chatbot-respond-relevant \end{tabular}} & \begin{tabular}[c]{@{}l@{}} \huge Responding relevantly and empathetically to \\ \huge user's relevant answers or questions \end{tabular} & \begin{tabular}[c]{@{}l@{}}\huge \textit{"U: I lost my dog when I was eight."} \\ \huge \textit{"S: \textbf{ Thanks for sharing.}}\\ \huge \textit{\textbf{ I'm sorry you had to go through that.}"}\end{tabular}\\
\hline
\multicolumn{1}{c}{\begin{tabular}[c]{@{}l@{}}\huge chatbot-acknowledge \end{tabular}} & \begin{tabular}[c]{@{}l@{}} \huge  Accepting and admitting user's inputs \end{tabular} & \begin{tabular}[c]{@{}l@{}}\huge \textit{"U: I don't like rollercoasters. "} \\ \huge \textit{"S: \textbf{Okay.} "}\end{tabular}\\
\hline
\multicolumn{1}{c}{\begin{tabular}[c]{@{}l@{}}\huge chatbot-ask-followup \end{tabular}} & \begin{tabular}[c]{@{}l@{}} \huge Asking follow-up questions based on \\ \huge recognized user intents \end{tabular} & \begin{tabular}[c]{@{}l@{}}\huge \textit{"U: Least favorite was physics. "} \\ \huge \textit{"S: \textbf{Do you mind saying a bit more?} "}\end{tabular}\\
\hline

\multicolumn{1}{c}{\begin{tabular}[c]{@{}l@{}}\huge chatbot-handle-user- \\ \huge -question-irrelevant \end{tabular}} & \begin{tabular}[c]{@{}l@{}} \huge Handling user's questions about topics \\ \huge different from the current interview topic     \end{tabular} & \begin{tabular}[c]{@{}l@{}} \huge \textit{"U: How are you feeling?"}\\ \huge \textit{"S: \textbf{"Thank you for asking. I feel great!"}}\end{tabular} \\
\hline
\multicolumn{1}{c}{\begin{tabular}[c]{@{}l@{}}\huge chatbot-handle-user-excuses \end{tabular}} & \begin{tabular}[c]{@{}l@{}} \huge Handling user's excuses to \\ \huge avoid answering the question/digression    \end{tabular} & \begin{tabular}[c]{@{}l@{}} \huge \textit{"S: "What were the worst parts of your childhood? "} \\ \huge \textit{"U: This is personal. "}\\ \huge \textit{"S: "\textbf{I understand. We'll continue then.}"}\end{tabular} \\
\hline
\multicolumn{1}{c}{\begin{tabular}[c]{@{}l@{}}\huge chatbot-handle-user-request \end{tabular}} & \begin{tabular}[c]{@{}l@{}} \huge Handling user's various requests  \end{tabular} & \begin{tabular}[c]{@{}l@{}} \huge \textit{"U: Tell me a joke. "}\\ \huge \textit{"S: \textbf{Why was the math book sad?}} \\ \huge \textit{\textbf{Because it had too many problems..}}"\end{tabular} \\
\hline
\multicolumn{1}{c}{\begin{tabular}[c]{@{}l@{}}\huge chatbot-handle-user-command \end{tabular}} & \begin{tabular}[c]{@{}l@{}} \huge Handling user's commands \\ \huge on managing the chat-flow    \end{tabular} & \begin{tabular}[c]{@{}l@{}} \huge \textit{"U: I want to skip the current questions. "}\\ \huge \textit{"S: \textbf{That's okay. Let's move on then.}"}\end{tabular} \\
\hline

\multicolumn{1}{c}{\begin{tabular}[c]{@{}l@{}}\huge chatbot-echo-user-\\ \huge -respond-irrelevant \end{tabular}} & \begin{tabular}[c]{@{}l@{}} \huge Responding to user's irrelevant responses \\ \huge relevantly and empathetically   \end{tabular} & \begin{tabular}[c]{@{}l@{}} \huge \textit{"S: What do you do now for a living?"} \\ \huge \textit{"U: I felt lonely sometimes."}\\ \huge \textit{"S: \textbf{If you need urgent help, please call 911 or}} \\ \huge \textit{\textbf{your doctor directly. I'd love to cheer you up if I could. }"}\end{tabular} \\
\hline
\multicolumn{1}{c}{\begin{tabular}[c]{@{}l@{}}\huge chatbot-handle-user-complain \end{tabular}} & \begin{tabular}[c]{@{}l@{}} \huge Handling user's complaints \end{tabular} & \begin{tabular}[c]{@{}l@{}} \huge \textit{"U: "You didn't listen. I just answered it. "} \\ \huge \textit{"S: \textbf{Sorry, I must have missed it."}}\end{tabular}\\
\hline
\multicolumn{1}{c}{\begin{tabular}[c]{@{}l@{}}\huge chatbot-social-obligations \end{tabular}} & \begin{tabular}[c]{@{}l@{}} \huge Handling user's acknowledging or\\ \huge social obligation inputs \end{tabular} & \begin{tabular}[c]{@{}l@{}} \huge \textit{"U: Thank you. "}\\ \huge \textit{"S: "\textbf{You're most welcome, \{user's first name\}.}}\\ \end{tabular} \\
\hline
\multicolumn{1}{c}{\begin{tabular}[c]{@{}l@{}} \huge chatbot-respond-default-fallback \end{tabular}}  & \begin{tabular}[c]{@{}l@{}} \huge Not understanding user inputs and \\ \huge responding with default fallback messages \end{tabular} & \begin{tabular}[c]{@{}l@{}}\huge \textit{"S: \textbf{My bad, I didn't recognize your inputs.}} \\ \huge \textit{\textbf{ Let's try again."}} \end{tabular} \\
\hline
\multicolumn{1}{c}{ \begin{tabular}[c]{@{}l@{}}\huge chatbot-repeat \end{tabular} }& \begin{tabular}[c]{@{}l@{}} \huge Not understanding user's answers \\ \huge and repeat the same utterance again \end{tabular} & \begin{tabular}[c]{@{}l@{}}\huge \textit{"S: What things frighten you now?"} \\ \huge \textit{"U: nothing."} \\ \huge \textit{"S: \textbf{What things frighten you now?"}} \end{tabular} \\
\hline
\multicolumn{1}{c}{ \begin{tabular}[c]{@{}l@{}}\huge chatbot-handle-gibberish \end{tabular} }& \begin{tabular}[c]{@{}l@{}} \huge Handling user's gibberish \end{tabular} & \begin{tabular}[c]{@{}l@{}}\huge \textit{"U:blea blahe}" \\ \huge  \textit{"S:\textbf{ Sorry I didn't understand. Please use English.}"}\end{tabular} \\
\hline
\multicolumn{1}{c}{ \begin{tabular}[c]{@{}l@{}}\huge chatbot-other \end{tabular} }& \begin{tabular}[c]{@{}l@{}} \huge Chatbot responses do not belong to \\ \huge any of the categories above \end{tabular} & \begin{tabular}[c]{@{}l@{}}\huge  \textit{"S:\textbf{ Sorry I got disconnected. Let's continue.}"}\end{tabular} \\
\hline
\multicolumn{2}{c}{ \begin{tabular}[c]{@{}l@{}}\huge * Note: "S" denotes the chatbot system while "U" denotes the user. \\ \huge Examples are for demostration purpose only, not necessarily from the original transcripts.\end{tabular}}
\end{tabular}}
  
~\label{tb:chatbottaxonomy}
\end{table*}

\subsection{Computational Framework for Evaluating and Explaining Chatbot Response Inappropriateness}

Our computational framework has two primary goals: 1) to provide a modeling workflow that enables non-technical designers to automatically detect and understand inappropriate chatbot responses; 2) to provide example-based explanations that facilitate a better understanding of chatbot inappropriateness. The framework addresses the problem of chatbot response appropriateness by formulating it as a three-class classification problem that distinguishes between appropriate, inappropriate, and neutral responses. 
Building upon prior research in DA characterization \cite{core1997coding, stolcke2000dialogue, xu2018towards, montenegro2019dialogue, wu2021semantic}, discourse theories \cite{grosz1986attention}, and DA classification \cite{raheja-tetreault-2019-dialogue}, we have developed a three-step modeling workflow for our proposed framework. This workflow consists of context-aware DA characterization, feature identification and automatic extraction with the characterized DAs, and modeling and explaining. An overview of our framework can be found in Figure~\ref{fig:framework}. In the following sections, we provide a detailed description of the framework and illustrate it with the interview chatbot dataset as a case study.

\subsubsection{Context-Aware DA Characterization in Human-Chatbot Interactions}
DA characterization is to model a single utterance in a dialogue with functional tags which represent the communicative intentions behind it. Since the same utterance can reflect different intentions due to different contexts, determining the DA category of one utterance requires context-aware modeling based on the preceding and following context \cite{raheja-tetreault-2019-dialogue}. 
With this in mind, the first step of our framework is an open coding process \cite{holton2007coding} to investigate the DAs that are frequently associated with users and chatbots in different conversational contexts in the dataset. Specifically with our dataset, we analyzed a subset of the dataset consisting of dialogues associated with the four interview topics presented in Table~\ref{tb:topictable}. To achieve this, we randomly selected 100 dialogues belonging to each interview topic, and an expert evaluator manually annotated each utterance in the subset with a label that best describes its DAs considering the contexts. 
After analyzing the occurrences of DAs and grouping similar DAs into categories, we identified 12 user DAs and 14 chatbot DAs. Tables~\ref{tb:usertaxonomy} and~\ref{tb:chatbottaxonomy} present the DA characterization for users and chatbots in human-chatbot interactions, respectively.

\begin{figure}[htb]
     \centering
         \includegraphics[width=0.5\textwidth, height=0.22\textheight]{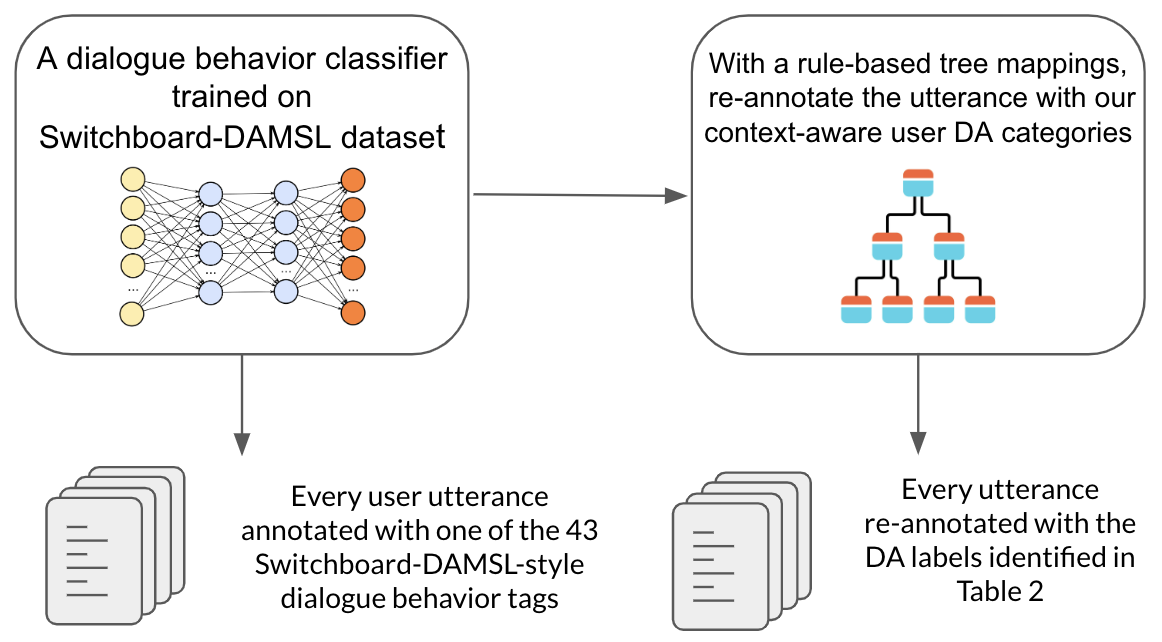}
         \caption{The cascading method we used to realize DA auto-annotation for an utterance}
         \Description{This figure shows the cascading method we used to realize DA auto-annotation for an utterance. The details of the method is described in section 2.2.1.}
      \label{fig:cascadingmethod}
\end{figure} 
 
\subsubsection{Identifying and Automatically Extracting Features with DA Characterization}
Drawing on previous discourse analysis theories by Grosz and Sidner \cite{grosz1986attention}, our framework identifies key features in human-chatbot dialogues from three different levels: \textit{linguistic level}, \textit{intentional level} and \textit{attentional level}. 
At the \textit{linguistic level}, our framework identifies specific linguistic markers, such as words or phrases, that contribute to the \textit{shallow discourse structure} \cite{core1997coding, jurafsky1997switchboard, stolcke2000dialogue}. In our case study with interview chatbots, we utilized the interview topic's and target chatbot responses' unigram bag of words as \textit{linguistic-level} features. Meanwhile, the \textit{intentional level} captures the utterance-level DAs. We thus encoded the \textit{intentional-level} features through one-hot categories of various dialogue components including the target chatbot utterance, all previous chatbot utterances, all following chatbot utterances, the most recent user utterance before the target, all previous user utterances, the next user utterance after the target, all following user utterances. The \textit{attentional level} models the dynamic focus of attention as the dialogue unfolds and the relationships between utterances, contributing to the \textit{deep discourse structure} \cite{core1997coding, jurafsky1997switchboard, stolcke2000dialogue}. For simplicity, we utilized the user-chatbot exchange DA pairs and ordinal index of the target chatbot response to describe the \textit{attentional-level} features.

Although most of the identified features mentioned above can be extracted computationally, the categorization of utterance-level user DAs still requires additional annotation efforts. To automate this process, we propose a two-stage cascading method for auto-annotating each user utterance's DAs (Figure~\ref{fig:cascadingmethod}). The first stage employs a dialogue behavior classifier that is trained on a large-scale open-sourced dataset, specifically the Switchboard-DAMSL dataset\cite{jurafsky1997switchboard, stolcke2000dialogue}, to assign Switchboard-DAMSL-style dialogue behavior tags to the utterances (pre-annotation). The Switchboard-DAMSL dataset contains a tag set of 43 mutually exclusive dialogue behaviors with the intention of building better language models for conversations. We directly utilized the dialogue behavior classifier trained by Raheja and Tereault \cite{raheja-tetreault-2019-dialogue} in this stage. Subsequently, in the second stage, these pre-annotated utterances are re-annotated automatically, following a rule-based tree mapping 
between the 43 Switchboard-DAMSL-style dialogue behavior tags and our characterized context-aware user DA categories from Table~\ref{tb:usertaxonomy}. An expert evaluator formulated the rule-based tree mapping manually following an open coding process \cite{holton2007coding}. The mapping was created with the same subset of the dataset used during the context-aware DA characterization phase.

\subsubsection{Modeling and Explaining with DA Characterization}
The framework then utilizes the extracted features to train classifiers for automatic detection of chatbot response appropriateness. To democratize the evaluation process, the framework opts to employ simple and interpretable machine learning models such as random forest (RF), instead of deep learning models that are resource-intensive and opaque. Following the common practice, the framework utilizes grid search to select the hyperparameters for the model. For the evaluation, the framework assesses model performance using four standard performance metrics, namely precision, recall, F1, and accuracy. To adjust for class imbalance, the framework weights all the metrics by the number of samples in each class when reporting the overall model performance. With the auto-annotated utterance DAs and detection results, the framework further examines the mismatch between the contextual utterances' DAs and the target chatbot response DAs to explain the inappropriateness. We present the results of modeling and explaining in the following section.

\section{Results}
\subsection{Modeling} Following the modeling practice in the framework, we trained an RF model to detect and explain the chatbot response inappropriateness. Using our interview chatbot dataset, we allocated 80\% of the data to the training set and the remaining 20\% to the test set. Our framework has achieved an accuracy of 91.0\%. To further validate the effectiveness of our framework, we compared our RF model with a baseline model on the same dataset. The baseline is a RoBERTa classifier fine-tuned on our dataset, which is a complex deep learning model that has demonstrated top performance in many natural language processing (NLP) tasks \cite{liu2019roberta}. We used 10\% of the dialogues in the training set as the development set for hyperparameter selection.  Table~\ref{tb:results} shows the performance of the two models. 

\begin{table*}
\renewcommand\arraystretch{1.2}
\caption{Evaluation Results of Chatbot Response Inappropriateness}
\begin{tabular}{c|l|l|l|l|l|l|l}
\hline
Model                                                                                                                 & \multicolumn{1}{c|}{Class} & \multicolumn{1}{c|}{Precision} & \multicolumn{1}{c|}{Recall} & \multicolumn{1}{c|}{F1} & \multicolumn{1}{c|}{\begin{tabular}[c]{@{}c@{}} Accuracy\end{tabular}} & \multicolumn{1}{c|}{\begin{tabular}[c]{@{}c@{}}Required \\ Computational Resources\end{tabular}}                                                   & \multicolumn{1}{c}{Explainability} \\ \hline

                                                                                                                      & Inappropriate              & 0.810                          & 0.781                       & 0.795                   &                                                                                   &                                                                                                                                                    &                                    \\ \cline{2-5}
                                                                                                                      & Neutral                    & 0.942                          & 0.931                       & 0.936                   &                                                                                   &                                                                                                                                                    &                                    \\ \cline{2-5}
\multirow{-3}{*}{\begin{tabular}[c]{@{}c@{}}Fine-tuned RoBERTa\\ (Baseline)\end{tabular}}                                & Appropriate                & 0.909                          & 0.965                       & 0.936                   & \multirow{-3}{*}{0.906}                                                  & \multirow{-3}{*}{\begin{tabular}[c]{@{}l@{}}High training/prediction \\ \& High storage efficiency \redcross \end{tabular}}                      & \multirow{-3}{*}{Low \redcross}            \\ \hline
                                                                                                                      & Inappropriate              & 0.789                          & 0.819                       & 0.804                   &                                                                                   &                                                                                                                                                    &                                    \\ \cline{2-5}
                                                                                                                      & Neutral                    & 0.959                          & 0.928                       & 0.943                   &                                                                                   &                                                                                                                                                    &                                    \\ \cline{2-5}
\multirow{-3}{*}{\begin{tabular}[c]{@{}c@{}}RF with \\ our proposed framework\end{tabular}}                           & Appropriate                & 0.916                          & 0.955                       & 0.935                   & \multirow{-3}{*}{\textbf{0.910}}                                                           & \multirow{-3}{*}{\begin{tabular}[c]{@{}l@{}}\textbf{Low training/prediction} \\ \textbf{\& Low storage efficiency} \greencheck \end{tabular}}                        & \multirow{-3}{*}{\textbf{High} \greencheck }           \\ \hline
\end{tabular}
~\label{tb:results}
\end{table*}

\subsection{Explaining} 
\begin{figure}[htb]
     \centering
         \includegraphics[width=0.5\textwidth, height=0.16\textheight]{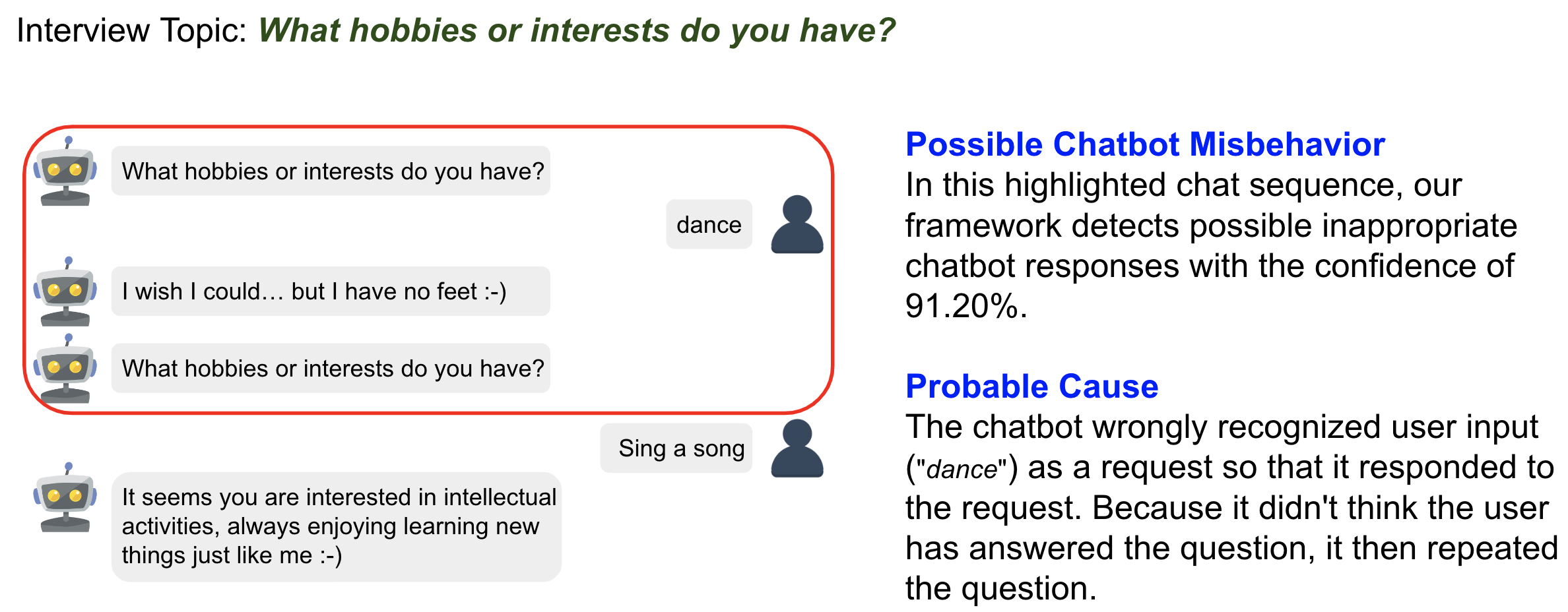}
         \caption{An example-based explanation generated by our framework through examining the mismatch between the DAs behind the inappropriate chatbot response and the DAs behind the contextual utterances.}
         \Description{This figure shows an example-based explanation generated by our framework examining the mismatch between the DAs behind the inappropriate chatbot response and the DAs behind the contextual utterances. }
      \label{fig:prototype}
\end{figure}

Our detection model enables us to generate example-based explanations by examining the mismatch between the DAs behind the inappropriate chatbot response and the DAs behind the contextual utterances. These explanations remind chatbot designers of the probable causes of inappropriate responses generated by their chatbot designs. For instance, in Figure~\ref{fig:prototype}, we can observe that the chatbot incorrectly recognized the user's input (\textit{"dance"}) as a "\textit{user-request}" (from Table~\ref{tb:usertaxonomy}) and responded accordingly ("\textit{chatbot-handle-user-request}" from Table~\ref{tb:chatbottaxonomy}). However, since the chatbot did not perceive the user's input as an answer to its question, it repeated the question ("\textit{chatbot-repeat}" from Table~\ref{tb:chatbottaxonomy}), resulting in an inappropriate response.

\section{Discussion}
We discovered that our model achieved comparable performance to RoBERTa (91.0\% vs. 90.6\%) while utilizing fewer computational resources and offering greater model simplicity, resulting in higher interpretability. This comparison highlights the effectiveness of our proposed features and the potential of incorporating DA modeling in detecting inappropriate chatbot responses. During our experiments, RoBERTa required significantly more computational resources for both training and prediction than our model. It took us 1 hour and 52 minutes to fine-tune RoBERTa (4 epochs) on single NVIDIA Tesla K80 GPU and 1 minute and 49 seconds to make predictions, while our model required no specialized hardware and only needed 11.6 seconds to finish training and less than 0.5 seconds to make predictions. Our model's storage efficiency is also much higher than RoBERTa since RF's storage efficiency is proportional to the number of decision trees (500) in the ensemble and the maximum depth of each tree (45), whereas RoBERTa has hundreds of millions of parameters (123 million) that need to be stored. 
Additionally, the RF model's simpler architecture enables us to provide easy-to-interpret features and decision paths associated with specific chatbot responses. In contrast, RoBERTa is often considered a black box \cite{devlin2018bert}, making it difficult to interpret how it makes predictions. Benefiting from such high explainability, our framework offers example-based explanations with corresponding DA tags and contexts to guide chatbot designers in the next design iteration. With the explanations, chatbot designers can better understand the probable causes and devise appropriate strategies to fix any inappropriate responses that fall within the same mismatched DA categories. 
The comparison between our model and RoBERTa demonstrates that our framework can democratize chatbot inappropriateness debugging to non-technical users in terms of requiring fewer computational resources and offering higher explainability while maintaining relatively good detection performance.

\section{Conclusion and Future Works}
Our findings indicate the feasibility and effectiveness of our proposed computational framework in evaluating and explaining chatbot inappropriateness. By incorporating DA modeling with just a simple RF model, our framework achieved comparable performance to top deep learning models while offering higher explainability and requiring fewer computational resources. In actual practice, our computational model can help chatbot designers identify the inappropriate responses from the pilot data and make corresponding revisions in further design iterations.  These features make our framework an effective tool for non-technical chatbot designers to iteratively evaluate and improve their designs, which greatly democratizes the chatbot debugging process. However, we acknowledge some challenges and opportunities for future studies, such as:
\begin{itemize}
  \item Exploring Framework Generalization Capability: While our results demonstrate promising performance in the context of interview chatbots, the generalizability of these findings to chatbots in diverse domains and with respect to other types of misbehaviors, such as toxic behaviors \cite{han2019evaluating}, remains uncertain. Additionally, it is worth investigating the necessary adaptations required to enhance the framework's applicability and generalizability.
  \item Interviewing Chatbot Designers: Since the target audience of our framework is non-technical chatbot designers, it is essential to test its usability and gather feedback from designers themselves to improve its practicality and effectiveness.
  \item Enhancing the Framework's Design-Assisting Capability: In its current stage, our framework provides example-based explanations of inappropriate chatbot responses with characterized DA tags and contexts. Inspired by previous work \cite{han2021designing}, we aim to provide more actionable design suggestions based on these examples to improve the democratization level of chatbot debugging. 
\end{itemize}

\bibliographystyle{ACM-Reference-Format}
\bibliography{acmart}

\end{document}